\documentclass[aps,prb,twocolumn,groupedaddress,floatfix,superscriptaddress]{revtex4-1}
\usepackage{epsfig}
\usepackage{amsmath,amssymb,graphicx,ulem}
\usepackage{physics}
\usepackage[dvipsnames,usenames]{xcolor}

\begin{document}

\title{Pairing and superconductivity in quasi one-dimensional flat
  band systems: Creutz and sawtooth lattices}

\author{Si Min Chan}

\affiliation{Centre for Quantum Technologies, National University of
  Singapore, 2 Science Drive 3, 117542 Singapore}

\affiliation{Department of Physics, National University of Singapore,
  2 Science Drive 3, 117542 Singapore}

\author{B. Gr\'emaud}

\affiliation{Aix Marseille Univ, Universit\'e de Toulon, CNRS,
  CPT, Marseille, France}
 
\author{G. G. Batrouni}

\affiliation{Universit\'e C\^ote d'Azur, INPHYNI, CNRS, 06103 Nice,
  France}

\affiliation{Centre for Quantum Technologies, National University of
  Singapore, 2 Science Drive 3, 117542 Singapore}

\affiliation{Department of Physics, National University of Singapore,
  2 Science Drive 3, 117542 Singapore}

\affiliation{Beijing Computational Science Research Center, Beijing
  100193, China}

\begin{abstract}
  We study the pairing and superconducting properties of the
  attractive Hubbard model in two quasi one-dimensional topological
  lattices, the Creutz and sawtooth lattices, which share two peculiar
  properties: each of their band structures exhibits a flat band with
  a non-trivial winding number. The difference, however, is that only
  the Creutz lattice is genuinely topological, owing to a chiral
  (sub-lattice) symmetry, resulting in a quantized winding number and
  zero energy edge modes for open boundary conditions.  We use
  multi-band mean field and exact density matrix renormalization group
  in our work. Our three main results are: (a) For both lattice
  systems, the superconducting weight, $D_s$, is linear in the
  coupling strength, $U$, for low values of $U$; (b) for small $U$,
  $D_s$ is proportional to the quantum metric for the Creutz system
  but not for the sawtooth system because its sublattices are not
  equivalent. We have therefore extended the approach to this more
  complex situation and found a excellent agreement with the numerical
  results; (c) at moderate and large $U$, conventional BCS mean field
  is no longer appropriate for such systems with inequivalent
  sublattices. We show that, for a wide range of densities and
  coupling strengths, these systems are very well described by a full
  multi-band mean field method where the pairing parameters and the
  local particle densities on the inequivalent sublattices are
  variational mean field parameters.
\end{abstract}

\maketitle 

\section{Introduction}
Systems with dispersionless (flat) bands have been the focus of much
interest due to the wide range of exotic quantum phases they can
exhibit. In such models, even infinitesimal interactions are much
larger than the band width, resulting in strongly correlated
physics. This was argued \cite{khodel90,kopnin11,heikkila11} to lead
to much higher critical temperatures for the transition to a
superconducting (SC) phase. Interest in such systems intensified
dramatically with the discovery of nonconventional SC in bi-layer
graphene \cite{cao18} twisted at a ``magic'' angle which causes a flat
band to appear in the band structure of the system. The energy of a
particle in a flat band is independent of its momentum which results
in very high degeneracy and the localization of the particle on a few
sites due to destructive quantum interference
\cite{schulenburg02,zhitomirsky04,derzhko09}. For weak attractive
interaction (much smaller than the gap between the flat band and other
bands), it was argued, using BCS mean field theory (MF), that the SC
weight is linearly dependent on the interaction strength
\cite{peotta15,tovmasyan16,liang17,verma21} and proportional to the
quantum metric, thereby linking the SC weight to the topological
properties of the non-interacting
system~\cite{provost80,berry89,Chiu2016}. This was confirmed for the
Creutz lattice with exact numerical calculations
\cite{mondaini18,tovmasyan18} and for a two-dimensional model using
determinant quantum Monte Carlo\cite{chowdhury20}. The main assumption
behind this prediction is that the gap function, $\Delta$ (see below),
is uniform, which is the case for a class of flat band systems like
the Creutz lattice. However, there are other systems of theoretical
and experimental interest, such as the sawtooth lattice
\cite{pyykkonen21}, where this assumption is not valid as the gap
function and site density are sublattice dependent. It is thus
important to identify how these predictions of the dependence on the
coupling and quantum metric change. In fact, as we show below, the BCS
MF treatment itself needs to be reexamined and modified. In
Ref.\cite{mondaini18} it was shown that in the large $U$ limit, the
attractive Hubbard model on the Creutz lattice is well represented by
an effective hard core boson model in a non-flat band and with near
neighbor repulsive interaction\cite{emery76,micnas90}.  With the low
and high $U$ limits studied, an interesting question arises: Can one
calculate the properties of the system in the intermediate coupling
regime where $U$ is comparable to the gap between the bands, and
where, consequently, there is strong band mixing?
  
Here, we address these issues by studying superconductivity in the
attractive Hubbard model with balanced populations in two quasi
one-dimensional lattices which exhibit a flat band in their ground
state; the Creutz \cite{creutz99} and sawtooth \cite{sen96,nakamura96}
lattices. Our main results are as follows. Using both multi-band BCS MF
and exact density matrix renormalization group (DMRG) calculations, we
determine the SC weight and the gap functions, and show excellent
agreement over a wide range of particle densities and couplings. This
is rather remarkable because at very low interaction, the flat band
(and lattice topology) determine the physics; at intermediate
coupling, $\abs{U}$ of the order of the gap between the bands, strong
inter-band mixing determines the physics; and at very strong coupling,
the physics is determined by an effective model of hard core bosons
with near neighbor repulsive interaction in a non-flat
band\cite{emery76,micnas90}. In spite of the different mechanisms
dominant in these three regimes, the multi-band mean field reproduces
the physics faithfully. Another main result is that for the sawtooth
lattice, where the two sublattices are not equivalent, it is necessary
to do a multi-band MF in both the sublattice-dependent densities and
gap functions. We show that the order parameter, {\it i.e.}  the pair
wavefunction, jumps to a large finite value for any nonzero
attraction in flat band systems. This is in stark contrast with
dispersive bands where the order parameter is exponentially small for
weak attraction.

\section{Model and methods} 
\begin{figure}[ht]
  \includegraphics[width=8cm]{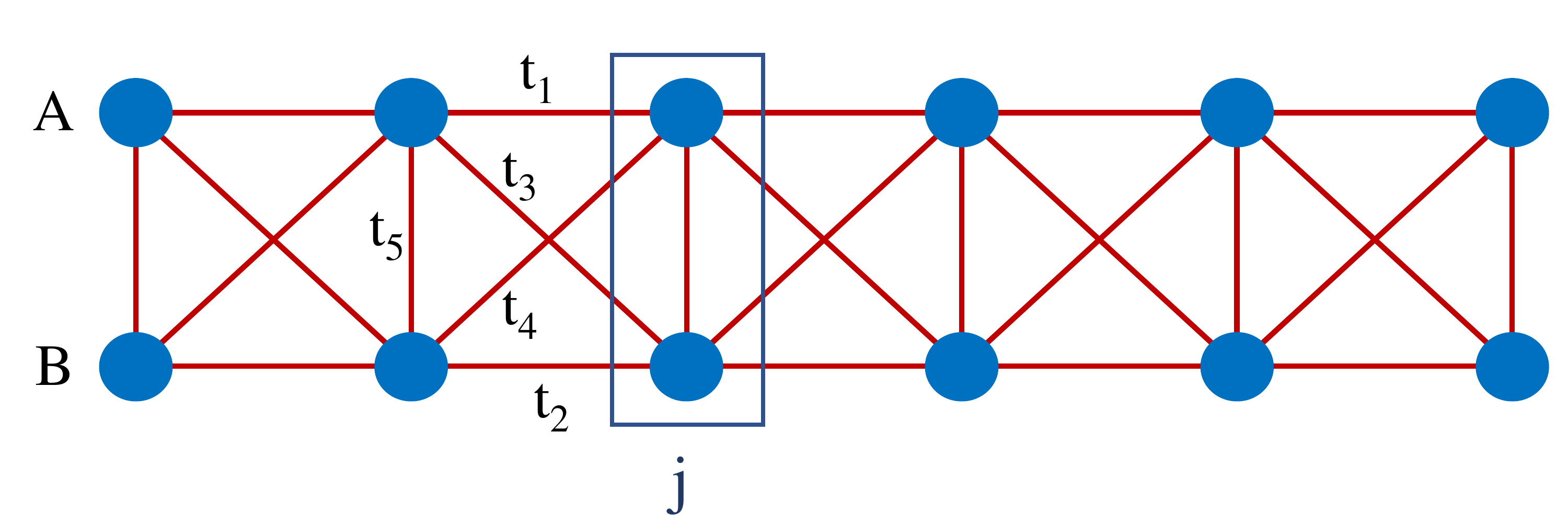}
 \caption{(Color online) The flat band Creutz lattice is given by
   $t_1=t_2^*=it$, $t_3=t_4=t$ and $t_5=0$. The flat band sawtooth
   lattice is given by $t_1=t_4=0$, $t_2=t$, $t_3=t_5=\sqrt{2}t$. The
   normal ladder is given by $t_1=t_2=t_5=t$ and $t_3=t_4=0$. The
   lower (upper) chain is the $\alpha=$B ($\alpha=$A) leg; the (black)
   rectangle shows the unit cell with lattice coordinate $j$.}
 \label{lattice}
\end{figure}

We study the Hubbard model on a two-leg ladder governed by the
Hamiltonian,
\begin{align}
  H&=\sum_{i,j,\alpha,\beta,\sigma} (t_{ij}
  c^{\alpha\dagger}_{i\sigma}c^{\beta\phantom\dagger}_{j\sigma} +
  H.c.)  -U\sum_{i,\alpha} c^{\alpha\dagger}_{i\downarrow}
  c^{\alpha\dagger}_{i\uparrow} c^{\alpha\phantom\dagger}_{i\uparrow}
  c^{\alpha\phantom\dagger}_{i\downarrow}\nonumber\\
  &\phantom{=}- \mu\sum_{i,\alpha} n^\alpha_i,
  \label{hubham}
\end{align}
where $t_{ij}$ is the hopping parameter connecting lattice sites as
shown in Fig.\ref{lattice}, $c^{\alpha\,\phantom\dagger}_{i\sigma}$
($c^{\beta\,\dagger}_{i\sigma}$) destroys (creates) a fermion of spin
$\sigma$ on site $i$ on the $\alpha={\rm A,\, B}$ ($\beta={\rm A,\,
  B}$) leg of the ladder. The number operator is $n^{\alpha}_{i}
=n^{\alpha}_{i\downarrow} + n^{\alpha}_{i\uparrow}$, with
$n^{\alpha}_{i\sigma}= c^{\alpha\dagger}_{i
  \sigma}c^{\alpha\phantom\dagger}_{i \sigma}$, $\mu$ is the chemical
potential and $U>0$ is the (attractive) Hubbard interaction
parameter. The density, $\rho$, is defined as the total number of
particles divided by the number of sites.
\begin{figure}[ht]
  \includegraphics[width=8cm]{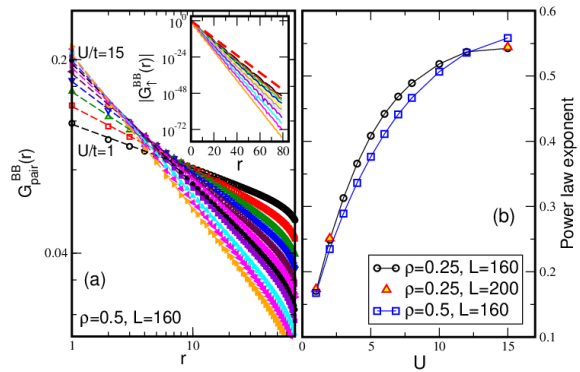}
 \caption{(Color online) DMRG results for the sawtooth lattice: (a)
   Pair Green function on the $B$ sublattice
   ($U=1,2,3,4,5,6,7,8,10,12,15$, $\rho=0.5$) exhibiting power law
   decay. The inset shows the exponential decay for the single
   particle Green function for the same cases. The thick (red) dashed
   line shows the analytically calculated $U\to 0$ limit giving a
   correlation length of $0.759$ lattice spacings (see text). (b) The
   exponents of the power law decay in (a) and also for the case of
   $\rho=0.25$. }
 \label{greens}
\end{figure}
\newline
Taking $t_1=it=t^*_2$, $t_3=t_4=t$, $t_5=0$, gives the Creutz lattice
with two flat bands at energies \cite{creutz99} $\pm 2t$, and a band
gap of $4t$; taking $t_1=t_4=0$, $t_2=t$, $t_5=t_3=\sqrt{2}t$ gives
the sawtooth lattice\cite{huber10} with one flat band at $-2t$ and one
dispersive band $\epsilon(k)=2t(1+{\rm cos}(k))$, and a band gap of
$2t$, with the lattice momentum $k$ an integer multiple of $2\pi/L$
where $L$ is the number of unit cells. In what follows we make these
flat band choices, our main goal being the study of pairing and the
resulting superconducting phases. Since in the Hamiltonian,
Eq.(\ref{hubham}), the attractive interaction is of the contact
on-site form, we expect on-site S-wave pairing to emerge, i.e. the
maximum of the pair wavefunction corresponds to both $\uparrow$ and
$\downarrow$ fermions being on the same site. Furthermore, the contact
interaction implies that, in a mean-field approach, only on-site
pairing terms are non-vanishing. A hallmark of pairing and pair
transport is that the pair (single particle) Green function decays as
a power (exponentially) with distance. These functions are given by,
\begin{align}
  G^{\alpha \beta}_\sigma(r)&=\langle c^\alpha_{j+r\sigma}
  c^{\beta\dagger}_{j\sigma}\rangle,
  \label{green}\\
  G_{pair}^{\alpha\beta}(r)&= \langle c^{\alpha}_{j+r\downarrow}
  c^{\alpha}_{j+r\uparrow}  c^{\beta\dagger}_{j\uparrow}
  c^{\beta\dagger}_{j\downarrow} \rangle.
  \label{pairgreen}
\end{align}
Another very important quantity characterizing the SC phase is the SC
weight, $D_s$, defined in one dimension by
\cite{kohn64,zotos90,shastry90,scalapino93,hayward95},
\begin{equation}
  D_s\equiv \pi L \frac{d^2 E_{GS}(\Phi)}{d
    \Phi^2}\bigg |_{\Phi=0}
  \label{ds}
\end{equation}
where $E_{GS}(\Phi)$ is the ground-state energy in the presence of a
phase twist $\Phi$ applied via the replacement $c_{j\sigma}^\alpha \to
{\rm e}^{i\phi j}c_{j\sigma}^\alpha$ with $\phi=\Phi/L$. Since only
near neighbor cells are connected, $H$ will depend on the phase
gradient, $\phi$, which appears only in the hopping terms. 

We use MF and the ALPS \cite{alps} implementation of DMRG to calculate
$E_{GS}(\Phi)$ on a lattice with periodic boundary conditions (PBC)
which then yields $D_s$. Our PBC DMRG implementation is described in
Ref.\cite{mondaini18}, and allows us to reach up to $L=24$ unit
cells. $G^{\alpha \beta}_\sigma(r)$ and $G_{pair}^{\alpha\beta}(r)$
are calculated with DMRG with open boundary conditions (OBC) where
much larger sizes are achievable (up to $L=200$). When lattice sites
are equivalent, the BCS MF starts with the usual substitution,
$Uc^{\dagger}_{i\downarrow} c^{\dagger}_{i\uparrow}
c^{\phantom\dagger}_{i\uparrow} c^{\phantom\dagger}_{i\downarrow}
\rightarrow \Delta^{*} c^{\phantom\dagger}_{i\uparrow}
c^{\phantom\dagger}_{i\downarrow} + c^{\dagger}_{i\downarrow}
c^{\dagger}_{i\uparrow}\Delta -|\Delta|^2/U$.  Here, $\Delta \equiv
U\langle c^\alpha_{i\uparrow} c^\alpha_{i\downarrow}\rangle$ and
$\langle c^\alpha_{i\uparrow} c^\alpha_{i\downarrow}\rangle$ is the
site-independent order parameter, {\it i.e.} the wave function
describing pair condensation. In general, it is complex but can be
taken real when the lattice is uniform. However, when $U=0$, the
eigenstate of $H$ on the sawtooth lattice shows that the density on a
$B$ site is twice that on an $A$ site. This difference between $A$ and
$B$ sites is expected to persist for $U\neq 0$ and, therefore, be also
reflected in a difference between the order parameters on the two
sublattices. Consequently, the mean field calculation should allow for
distinct, unequal $\Delta^A$, $\Delta^B$, and for the local densities
on $A$ and $B$ sites to be treated as variational parameters. Using a
general quadratic trial Hamiltonian and applying the Gibbs-Bogoliubov
inequality \cite{kuzemsky15} (see Appendix \ref{appendix:mf}) we
obtain the MF Hamiltonian,
\begin{eqnarray}
  \nonumber
  H_{BCS}&=& \frac{LU}{2} + L\sum_{\alpha} \left [ (
  |\Delta^\alpha|^2+\theta^{z2 }_{\alpha} )/U\right ]\\
  \nonumber
  &&+ \sum_{i,j,\alpha,\beta,\sigma} (t_{ij} 
  c^{\alpha\dagger}_{i\sigma}c^{\beta\phantom\dagger}_{j\sigma} +
  H.c.)  -
  \sum_{i,\alpha}
  (\mu+\frac{U}{2})n^\alpha_i \\
    \nonumber
  &&-\sum_{i,\alpha}
    \theta^z_\alpha(n^{\alpha}_{i\uparrow}+n^{\alpha}_{i\downarrow}-1)\\
    &&-\sum_{i,\alpha}\left (
   \Delta^{\alpha
    *}c^\alpha_{i\uparrow} c^\alpha_{i \downarrow} + \Delta^\alpha
  c^{\alpha \dagger}_{i \downarrow} c^{\alpha \dagger}_{i \uparrow}
  \right ),
\label{bcsmftham}
\end{eqnarray}
where $\theta^z_{\alpha} \equiv U \langle n^\alpha_\downarrow +
n^\alpha_\uparrow -1 \rangle/2$. Fourier transforming, and defining
the Nambu spinor $\Psi_k^{\dagger} \equiv (c^{\rm A
  \dagger}_{k\uparrow}, c^{\rm B \dagger}_{k\uparrow}, c^{\rm A
  \phantom\dagger}_{-k\downarrow}, c^{\rm B
  \phantom\dagger}_{-k\downarrow})$, yields
\begin{eqnarray}
  \nonumber
  H_{BCS}(\Phi) &=& \sum_k \Psi^\dagger_k {\cal M}_k(\phi)
  \Psi_k\\
  &&+L\left(\frac{1}{U} \sum_{\alpha}
  (|\Delta^{\alpha}|^2+\theta_\alpha^{z2})-2 \mu{-\frac{U}{2}}\right),
  \label{bcsmfthamFT}
\end{eqnarray}
where ${\cal M}_k(\phi)$ is a $4\times 4$ Hermitian matrix (see
Appendix \ref{appendix:mf}) and where we now display explicitly the
phase twist $\Phi$ and its gradient $\phi$. It is clear that
$\theta^z_\alpha$ is real but, in general, $\Delta^\alpha$ is
complex. However, for lattices that are invariant under the
$\text{A}\leftrightarrow\text{B}$ exchange, e.g. ladder and Creutz
lattices, $\Delta$ is uniform and can be taken real for all
$\Phi$. For a time-reversal symmetric Hamiltonian, $\Delta^{\alpha}$
can also be taken real and this is the case for the sawtooth lattice
but only at $\Phi=0$. Finally, for lattices that are invariant under
the $\text{A}\leftrightarrow\text{B}$ exchange, the mean-field
parameters $\theta^z_{\alpha}$ also become homeogenous,
i.e. independent of $\alpha$, such that they can be absorbed in the
chemical potential $\mu$ leading back the usual simpler MF
Hamiltonian.

Diagonalizing ${\cal M}_k(\phi)$ yields the ground state energy,
$E_{GS}(\Phi)$, and allows us to solve the MF self-consistency
equations giving $\Delta^\alpha$ and $\theta^z_\alpha$ and then obtain
$D_s$. Apart from the additional MF parameters $\theta^{\alpha}$, our
MF approach is the same as the one used to describe the usual BCS-BEC
crossover, be it in lattices or in the bulk. However, the flat band
dramatically changes the behavior and nature of the pairing at low
interaction strength.

\section{Results and Discussion}
The power and exponential decays of the pair and single particle Green
functions on the Creutz and normal ladder lattices were presented in
Ref.\cite{mondaini18}. Here, we begin by establishing that the same
behavior occurs in the sawtooth lattice. Figure \ref{greens}(a) shows
the pair Green functions along the $B$ sublattice of the sawtooth
lattice for $U=1,2,3,4,5,6,7,8,10,12,15$ and $\rho=0.5$ exhibiting
clear power law decay. The corresponding exponents, and also the
exponents for the $\rho=0.25$ case, are shown in
Fig.\ref{greens}(b). It is seen that the exponents behave similar to
those for the Creutz lattice\cite{mondaini18} where they take smaller
values for smaller $U$, opposite to the behavior of the exponents on
the normal ladder\cite{mondaini18}. The inset of Fig.\ref{greens}(a)
shows the single particle Green functions for the same cases as in the
main panel and exhibits exponential decay over many decades, even for
small $U$.  This establishes that, as in the usual BCS theory, the
only transport in this system is via paired up and down
fermions~\cite{gremaud21}. However, for a flat band system, and in
sharp contrast with the standard BCS situation, the length scale
associated with the exponential decay of the single particle Green
functions does not diverge (exponentially) as the interaction strength
$U\rightarrow 0$. Instead it saturates to a finite value closely
related to the exponential decay of the Wannier function of the flat
band which is very fast. In addition, one can show that the pair
wavefunction (not to be confused with the pair Green function) is also
exponentially localized with a length scale of the same order as that
of the single particle Green function. For instance, at $\rho=0.5$,
one can show that both length scales have exactly the same value
$0.759$ lattice spacings, in perfect agreement with the results in the
inset of Fig.\ref{greens}(a). In other words, on conventional
lattices, the spatial extent of the pair is exponentially large for
small $U$ and decreases as $U$ increases, eventually becoming on-site
pairing. In the flat band systems considered here, the pairing is
essentially on-site for any value of $U$, even
infinitesimal. Therefore, the physics of this system is different from
that of the conventional BCS-BEC crossover.

\begin{figure}[ht]
  \includegraphics[width=8.5cm]{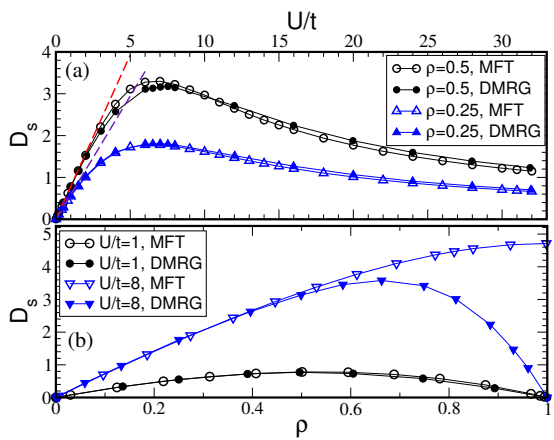}
 \caption{(Color online) Creutz lattice: (a) SC weight, $D_s$, {\it
     vs.} $U/t$ comparing BCS MF with DMRG\cite{mondaini18}.  Very
   good agreement is seen over a wide range of values. Dashed lines
   are the linear predictions \cite{tovmasyan16} for very small $U$
   (see text). (b) BCS MF agrees very well with DMRG over a wider
   range of $\rho$ for low $U$. For large $U$, the agreement is good
   at low densities. DMRG calculations were done on $L=16$ lattice
   with PBC.}
 \label{creutzDs}
\end{figure}

In Ref.\cite{mondaini18}, DMRG was used to calculate the SC weight,
$D_s$, for the Creutz lattice and shown to be linear for small $U$, as
predicted \cite{peotta15,tovmasyan16,liang17}. Using our multi-band
MF, Eq.(\ref{bcsmfthamFT}), we calculate $E_{GS}(\Phi)$ (see Appendix
\ref{appendix:mf}) and then $D_s$, Eq.(\ref{ds}), for the same
parameters in Ref.\cite{mondaini18}. We compare the MF and DMRG
results in Fig.\ref{creutzDs}. Figure \ref{creutzDs}(a) shows clearly
that the MF results are remarkably accurate for a very wide range of
$U$ values at the two densities studied. The dashed lines in
Fig.\ref{creutzDs}(a) are given by $D_s=\pi U\rho(1-\rho)$ derived
from the results of Ref.\cite{tovmasyan16} where it was assumed that
$U$ is much smaller than the gap between the bands. Consequently, at
these small values of $U$, the physics is dominated by the flat band,
and the states can be projected on it (see Appendix
\ref{appendix:ds}). This means that the upper band does not contribute
to the physics in this limit. With the added assumption that $\Delta$
is uniform and independent of the applied phase twist, it was
shown\cite{tovmasyan16} that for such very small values of the
coupling, $D_s$ is linear in $U$ and is proportional to the quantum
metric\cite{peotta15}, Eq.(\ref{appBquantmetric}). It is to be
emphasized that these approximations are valid only for $U$ much
smaller than the gap between the bands and, therefore, the physics at
very low $U$ is dominated by the flat band\cite{tovmasyan16} and the
topological properties of the lattice. The physics at intermediate $U$
(of the order of the interband gap) is dominated by very strong mixing
between the two bands, with both bands contributing significantly to
the properties of the system, such as $D_s$. At very strong $U$, the
physics is that of hard core bosons governed by a dispersive
Hamiltonian with near neighbor repulsion\cite{emery76,micnas90}. It is
remarkable that our multi-band MF calculation agrees with DMRG over
such a wide range of parameters despite the marked difference in
mechanism.

It is evident that the agreement between our MF and DMRG is better for
the lower density, $\rho=0.25$.  Figure \ref{creutzDs}(b) emphasizes
the dependence of the MF accuracy on the interplay of $U$ and
$\rho$. For $U=t$, MF gives a very good description for all fillings
up to half filling, $\rho=1$. For $U=8t$, MF is accurate only up to
around $\rho=0.5$. In particular, at half filling, $\rho=1$, the
system should be a band insulator (the lower band is full) but MF at
$U=8t$ yields a finite $D_s$.

\begin{figure}[ht]
  \includegraphics[width=8cm]{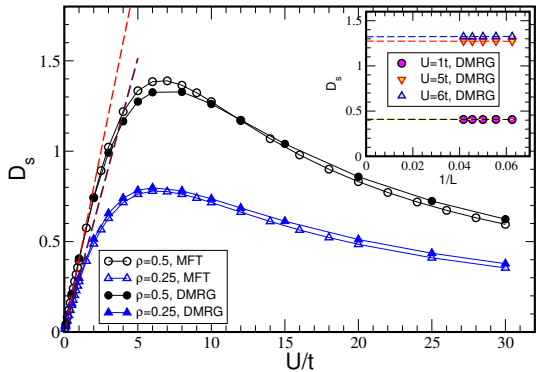}
 \caption{(Color online) SC weight, $D_s$, for the sawtooth lattice
   comparing MF with DMRG ($L=16$) showing excellent agreement. Dashed
   lines are linear fits at low $U$ where $D_s$ is predicted to be
   linear in $U$. See text for a discussion of the slopes. The inset
   shows $D_s$ {\it vs} $L^{-1}$ at $U=1,\,2,\,3$ for $\rho=0.5$ and
   illustrates that finite size effects are very small.}
 \label{sawtoothDs}
\end{figure}

\begin{figure}[h] 
 \centerline{\includegraphics[width=0.45\textwidth]{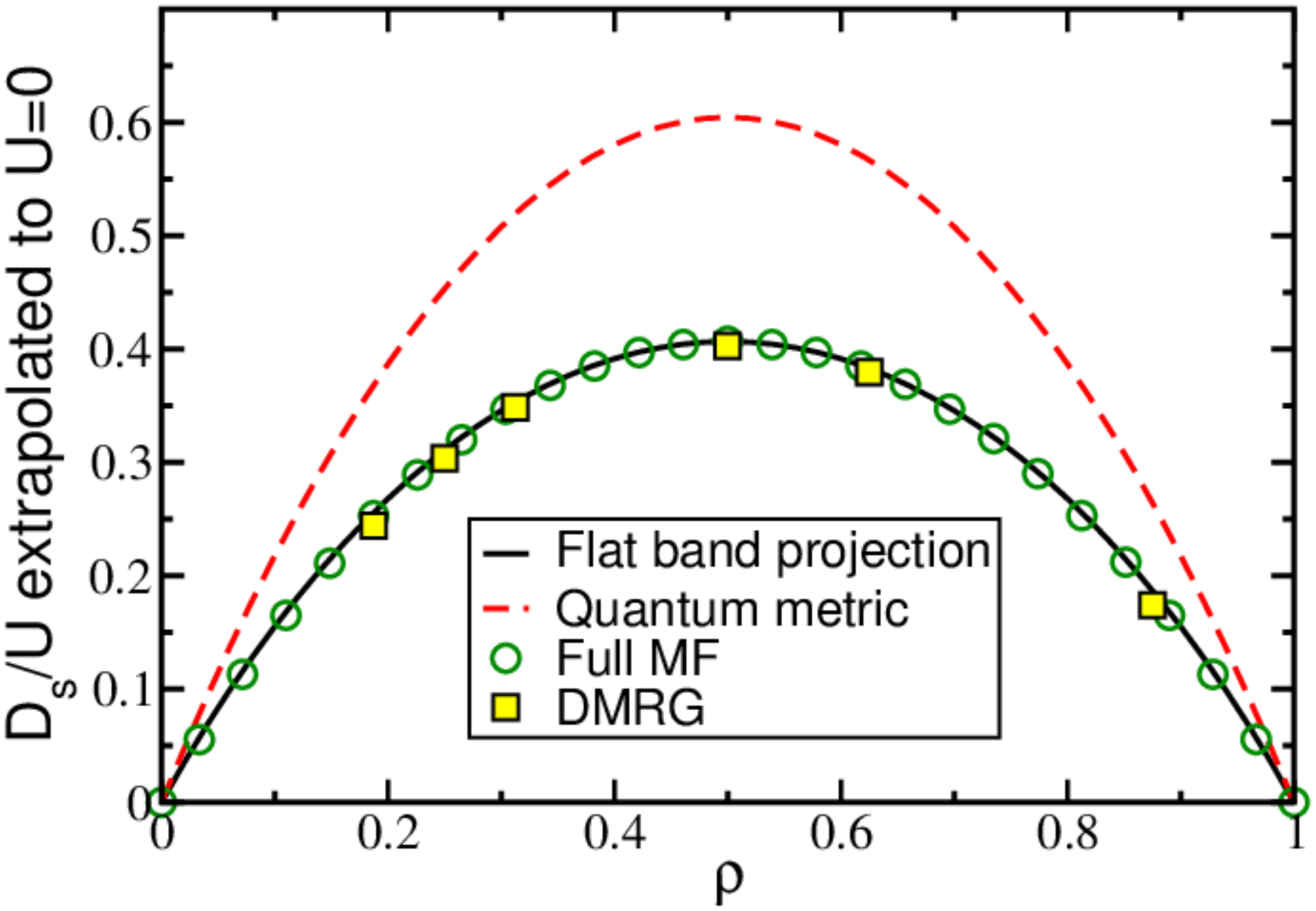}}
 \caption{\label{fig:ds_mfres} (Color online) Sawtooth lattice: SC
   weight $D_s/U$ extrapolated to $U=0$ as a function of the total
   density $\rho$ from the projection onto the flat band (black
   continuous line), compared with the full band computations (green
   circles) and with the results of ref.~\cite{peotta15}, which only
   include the quantum metric term (red dashed line). The yellow
   squares are the exact results obtained with DMRG.}
\end{figure}

With this agreement between MF and DMRG for the Creutz lattice, we now
study $D_s$ on the sawtooth lattice. We perform exact DMRG
calculations for the sawtooth lattice with PBC and sizes up to $L=24$
to calculate $D_s$ and compare with our extended BCS MF. Figure
\ref{sawtoothDs} shows $D_s$ {\it vs.} $U/t$ for two densities,
$\rho=0.5,\,0.25$. Similar to the Creutz, $D_s$ also increases
linearly at first, reaches a maximum and then decreases slowly. Figure
\ref{sawtoothDs} also shows that our multi-band MF values agree very
well with DMRG over the entire range of $U$ values we explored. We
emphasize that without the additional MF parameters
$\theta^z_{\alpha}$, agreement between MF and DMRG would be quite poor
for $D_s$ already at values of $U\approx 5t$ (see Appendix
\ref{appendix:mf}).  We note that the values of $D_s$ for sawtooth are
more than a factor of $2$ smaller than the corresponding values for
Creutz. The Creutz lattice offers a higher superfluid density for the
same particle density. Furthermore, the low $U$ behavior of $D_s$ is
linear in $U$, and the dashed lines in Fig.\ref{sawtoothDs} are fits
to the low $U$ linear parts of the curves; the slope for $\rho=0.5
(0.25)$ is $0.401(0.303)$. Using the results of
Ref.\cite{tovmasyan16}, which predict a slope proportional to the
quantum metric, $\mathcal{Q}=\frac{2}{3\sqrt{3}}$, we find slopes of
$0.6$ and $0.45$ for $\rho=0.5$ and $0.25$ respectively which do not
agree with the exact numerical values like they did for the Creutz
lattice. This disagreement is due to the fact that the $A$ and $B$
sublattices are not equivalent in the sawtooth case: $\Delta^A \neq
\Delta^B$ and $\rho^A \neq \rho^B$, and, consequently, the assumptions
in Refs.\cite{tovmasyan16,peotta15} which led to $D_s$ being
proportional to the quantum metric are no longer valid.  In fact, the
derivation in Ref.\cite{peotta15} not only assumes that, for all
values of the phase gradient $\phi$, $\Delta$ is the same for all
sites, but also that it has a vanishing first derivative with respect
to $\phi$, at $\phi=0$. It turns out that the latter assumption is, in
general, not valid as soon as $\Delta$ becomes sub-lattice dependent,
since the $U(1)$ symmetry only allows one to fix the global phase of
the mean-field parameters, leaving the possibility of a $\phi$
dependent relative phase between the $\Delta^{\alpha}$. We found that
when $\phi \neq 0$, the phase difference between $\Delta^A$ and
$\Delta^B$ for the sawtooth lattice is exactly equal to the phase
gradient $\phi$.  Nonetheless, by projecting carefully on the flat
band without making the above-mentioned simplifying assumptions, we
generalized the mean field computation of the slope of the superfluid
weight, see Appendix~\ref{appendix:ds}, as a function of the total
density $\rho$. The results are displayed in Fig.~\ref{fig:ds_mfres},
where we compare our analytical results for the slope at small $U$
with our full band computation and also to what one obtains by using
the results of Ref.~\cite{peotta15}. In addition we show the results
from our DMRG calculation. As can be seen clearly, the agreement
between all our approaches is excellent, but not with the results
predicted in Ref.~\cite{peotta15}, which only include the quantum
metric. In particular, our projection method gives the slope of $D_s$
at $\rho=0.5$ ($\rho=0.25$) to be $0.407$ ($0.311$) which compares
very well with the DMRG results $0.401$ ($0.303$).

\begin{figure}[ht]
  \includegraphics[width=8cm]{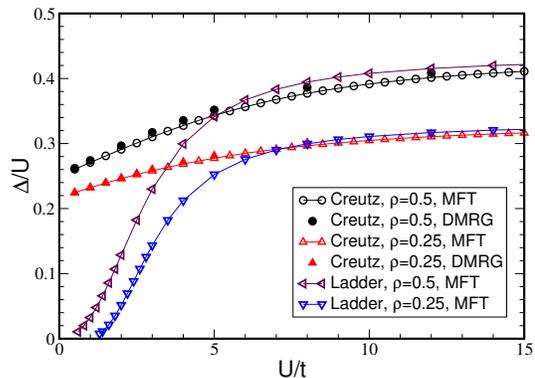}
 \caption{(Color online) The MF pairing parameter, $\Delta/U$, as a
   function of $U/t$ for the Creutz and normal ladder
   lattices. Agreement between DMRG and MF is excellent. }
 \label{deltasvsUcreutz}
\end{figure}

We elucidate further the agreement between our MF and exact DMRG
results by comparing the order parameter $\Delta^\alpha/U$ given by
the two methods. With MF, we calculate directly $\Delta^\alpha/U =
\langle c^\alpha_{i\uparrow} c^\alpha_{i\downarrow} \rangle $; with
DMRG, we use $\Delta^{\alpha *} \Delta^\alpha = \langle
c^{\alpha\dagger}_{i\downarrow} c^{\alpha\dagger}_{i\uparrow}
c^{\alpha\phantom\dagger}_{i\uparrow}
c^{\alpha\phantom\dagger}_{i\downarrow}\rangle - \langle
n^\alpha_{i\uparrow} \rangle \langle n^\alpha_{i\downarrow} \rangle$.
We show in Fig.\ref{deltasvsUcreutz} the MF and DMRG values for the
Creutz system and MF values for the normal ladder where for these
systems $\Delta^A=\Delta^B=\Delta$. Note that for the normal ladder
$\Delta/U$ vanishes exponentially as $U\to 0$ making accurate DMRG
determination not feasible because it requires exponentially large
systems. On the other hand, for the Creutz system $\Delta(U\to 0)/U$
is finite even though $\Delta(U=0) = 0$: Here, even for infinitesimal
attraction, the pairing parameter $\Delta/U$ acquires a large finite
value. We see in Fig.\ref{deltasvsUcreutz} excellent agreement between
the MF and DMRG values of $\Delta/U$ for a very wide range of $U$
values.  The difference in behavior of $\Delta/U$ between normal
ladder and Creutz lattices can be understood as follows: For a
dispersive band, it is well known that in the small $U/t$ limit,
pairing only occurs at the Fermi level and involves an (exponentially)
small fraction of the free fermions. Consequently, the pair density
itself becomes (exponentially) small. Furthermore, for small $U$, the
correlation between the paired electrons (the pair size) extends over
a very large distance. In contrast, there is no Fermi surface for a
flat band, and fermions at all momenta can form pairs leading to a
sizeable contribution to the pair density and yielding a finite value
even for arbitrarily small $U$. Additionally, even for infinitesimal
$U$, the system is strongly correlated because $U$ is much larger than
the width of the flat band. In this case, the pair size is essentially
on-site even for very small $U$. We emphasize that, even though the
order parameter does not vanish at small $U$, the superfluid weight,
$D_s$, and the charge gap (both controlled by $\Delta^{\alpha}$) do
vanish (linearly) with $U$.

\begin{figure}[ht]
  \includegraphics[width=9cm]{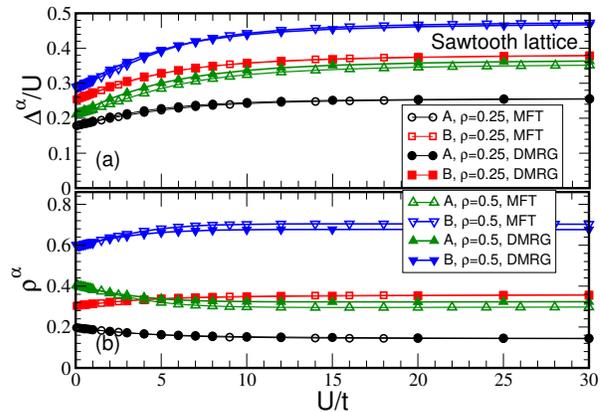}
 \caption{(Color online) Sawtooth lattice: (a) The MF pairing
   parameter, $\Delta^\alpha/U$, as a function of $U/t$; (b)
   $\rho^\alpha$ {\it vs} $U/t$.}
 \label{deltasvsUsaw}
\end{figure}
The sawtooth lattice MF and DMRG results are shown in
Fig.\ref{deltasvsUsaw}. Figure \ref{deltasvsUsaw}(a) shows
$\Delta^\alpha/U$ as a function of $U/t$ for two densities, $\rho=0.5$
and $0.25$. We see that the behavior as $U\to 0$ is qualitatively
similar to that of the Creutz lattice. However, here we observe the
imperative difference that, for the whole range of $U$, $\Delta^A\neq
\Delta^B$, and that our MF results are in excellent agreement with
DMRG. Furthermore, we see a similar pattern between $\rho^A$ and
$\rho^B$ in Fig.\ref{deltasvsUsaw}(b): for all $U$, their values
remain clearly different. These striking differences between the $A$
and $B$ sublattices cannot be obtained accurately with a simple BCS MF
calculation, which, at large $U$, inevitably leads to
$\Delta^A=\Delta^B$ and thereby in $\rho^A=\rho^B$ (see Appendix
\ref{appendix:mf}). To obtain correct behavior, we stress the
importance of treating the local densities as variational MF
parameters.

\section{Conclusions}
Using mean field and exact DMRG calculations, we studied the pairing
and superconducting properties of the attractive fermionic Hubbard
model in two flat band lattices with non-trivial winding numbers, the
Creutz and sawtooth lattices. The difference, however, is that only
the Creutz lattice is genuinely topological owing to a chiral
(sub-lattice) symmetry, resulting in a quantized winding number and
zero energy edge modes for open boundary conditions. On the contrary,
the lack of sublattice symmetry for the sawtooth lattice not only
results in a non-quantized winding number, but also causes the
densities and pairing parameters on the two sublattices to be
different and necessitates the use of a mean field method where the
local densities and pairing parameters are all MF parameters, as we
showed here. While mean field calculations may be expected to give
reasonably accurate results for weak coupling and low densities, we
show that our mean field describes the system remarkably well for a
very wide range of coupling values and densities. It only fails when
both the density and coupling attain high values,
Fig.\ref{creutzDs}(b). We emphasize one of our main results: since
$\Delta^A\neq \Delta^B$ for sawtooth, the superconducting weight,
$D_s$, is no longer simply proportional to the quantum metric for low
$U$ values, but is still linear in $U$. We calculated here the correct
and more general linear behavior (Appendix \ref{appendix:ds}). This is
expected to hold for any topological lattice where sublattices are not
equivalent. Finally, beyond static properties, our results emphasize
that for lattices with inequivalent sites, it is crucial to use the
extended mean field approach for time-dependent situations, such as AC
or DC Josephson effects~\cite{pyykkonen21}.

\underbar{\bf Acknowledgments:}
  S.M.C. is supported by a National University of Singapore
  President's Graduate Fellowship. The DMRG computations were
  performed with the resources of the National Supercomputing Centre,
  Singapore (www.nscc.sg). This research is supported by the National
  Research Foundation, Prime Minister's Office and the Ministry of
  Education (Singapore) under the Research Centres of Excellence
  programme.

\appendix

\section{Multi-band mean field method}
\label{appendix:mf}
The DMRG method we used for the exact calculation of $D_s$, and the
implementation of the periodic boundary condition are discussed in
Ref.\cite{mondaini18}. Here we discuss in some more detail the
multi-band MF method we use in this work.

We start with a general quadratic trial Hamiltonian,
\begin{eqnarray}
  \nonumber
  H_{trial} &=& \sum_{i,j,\alpha,\beta,\sigma} (t_{ij}
  c^{\alpha\dagger}_{i\sigma}c^{\beta\phantom\dagger}_{j\sigma} +
  H.c.) - \mu \sum_{i,\alpha} n^\alpha_i\\
  \nonumber
  &&- \sum_{i,\alpha} \left (
  \theta^\alpha_\downarrow n^\alpha_{i \uparrow} +
  \theta^\alpha_\uparrow n^\alpha_{i\downarrow}\right )\\
  && - \sum_{i,\alpha} \left ( \Delta^{\alpha *}
  c^\alpha_{i\uparrow} c^\alpha_{i\downarrow} + \Delta^\alpha
  c^{\alpha \dagger}_{i\downarrow} c^{\alpha \dagger}_{i\uparrow}
  \right ), 
  \label{trialham}
\end{eqnarray}
where $\theta^\alpha_\sigma$ and $\Delta^\alpha$ are the variational
MF parameters to be determined. The Gibbs-Bogoliubov
inequality\cite{kuzemsky15} gives an upper bound on the true free
energy, $F$, of the model governed by the Hamiltonian
Eq.(\ref{hubham}),
\begin{equation}
  F \leq {\rm Tr}[H W_{trial} ] + \frac{1}{\beta} {\rm Tr}[W_{trial}
    {\rm ln}W_{trial} ],
  \label{varfree}
\end{equation}
where the trial Boltzmann weight is given by,
\begin{equation}
  W_{trial} = \frac{1}{Z_{trial}} {\rm e}^{-\beta H_{trial}},
\end{equation}
and $Z_{trial} = {\rm Tr\,}{\rm e}^{-\beta H_{trial}} = {\rm e}^{-\beta
  F_{trial}}$. Substituting $W_{trial}$ in Eq.(\ref{varfree}) and
using Wick's theorm, we get,
\begin{eqnarray}
  F &\leq & -U\sum_{i,\alpha} \left ( \langle
  n^\alpha_{i\uparrow}\rangle \langle n^\alpha_{i\downarrow}\rangle +
  \langle c^{\alpha \dagger}_{i\downarrow}c^{\alpha
    \dagger}_{i\uparrow}\rangle \langle c^{\alpha
    \phantom\dagger}_{i\uparrow}c^{\alpha
    \phantom\dagger}_{i\downarrow} \rangle \right )\\
  \nonumber
  &&  + \sum_{i,\alpha} \left (
  \theta^\alpha_\downarrow \langle n^\alpha_{i\uparrow} \rangle +
  \theta^\alpha_\uparrow \langle n^\alpha_{i\downarrow} \rangle \right
  )\\
  \nonumber
  && + \sum_{i,\alpha} \left (\Delta^{\alpha *} \langle
  c^\alpha_{i\uparrow} c^\alpha_{i\downarrow} \rangle + \Delta^\alpha
  \langle c^{\alpha\dagger}_{i\downarrow} c^{\alpha
    \dagger}_{i\uparrow} \rangle \right ) + F_{trial}.
\end{eqnarray}
We minimize the right hand side with respect to the variational
parameters and obtain,
\begin{equation}
  \theta^\alpha_\uparrow = U\langle n^\alpha_{i\uparrow}
  \rangle,\,\,\,\theta^\alpha_\downarrow = U\langle n^\alpha_{i\downarrow}
  \rangle,\,\,\,\,\Delta^\alpha = U\langle c^\alpha_{i\uparrow}
  c^\alpha_{i\downarrow}\rangle.
  \label{mfparams}
\end{equation}
We substitute these expressions in Eq.(\ref{varfree}) to obtain the
optimized free energy,
\begin{eqnarray}
  \nonumber
  F &\leq& \frac{1}{U}\sum_{i,\alpha} (\theta^\alpha_\downarrow
  \theta^\alpha_\uparrow + |\Delta^\alpha|^2 ) -\frac{1}{\beta} {\rm
    ln\, Tr}\,{\rm e}^{-\beta H_{trial}}\\
  \nonumber
        &\leq & -\frac{1}{\beta} {\rm \ln\, Tr}\,{\rm e}^{-\beta \left (
      \frac{1}{U}\sum_{i,\alpha} (\theta^\alpha_\downarrow
      \theta^\alpha_\uparrow + |\Delta^\alpha|^2 ) + H_{trial} \right 
      ) } \\
  &\equiv& -\frac{1}{\beta} {\rm \ln\, Tr}\,{\rm e}^{-\beta H_{BCS}},
\end{eqnarray}
which defines $H_{BCS}$. Since we are dealing with systems with
balanced up and down populations, $\theta^\alpha_\uparrow =
\theta^\alpha_\downarrow$, we can instead define $\theta^z_\alpha
\equiv (\theta^\alpha_\uparrow + \theta^\alpha_\downarrow -
U)/2$. This allows us to rewrite,
\begin{eqnarray}
  \nonumber
  H_{BCS} &=& \frac{LU}{2} +L\sum_{\alpha} \left [ (
  |\Delta^\alpha|^2+\theta^{z2 }_{\alpha} )/U \right ]\\
  \nonumber
  && + \sum_{i,j,\alpha,\beta,\sigma} (t_{ij}
  c^{\alpha\dagger}_{i\sigma}c^{\beta\phantom\dagger}_{j\sigma} + 
  H.c.)  - 
  \sum_{i,\alpha}
  (\mu+\frac{U}{2})n^\alpha_i \\
  \nonumber
  &&-\sum_{i,\alpha} \left (\theta^z_\alpha(n^{\alpha}_{i\uparrow} +
  n^{\alpha}_{i\downarrow}-1 \right )\\
  && -\sum_{i,\alpha} \left ( \Delta^{\alpha
    *}c^\alpha_{i\uparrow} c^\alpha_{i \downarrow} + \Delta^\alpha
  c^{\alpha \dagger}_{i \downarrow} c^{\alpha \dagger}_{i \uparrow}
  \right ),
  \label{bcsmfthamSM}
\end{eqnarray}
which is Eq.(\ref{bcsmftham}). This can now be put in momentum space
via the Fourier transform,
\begin{equation}
  \chi_{r\sigma} = \frac{1}{\sqrt{L}}\sum_{k}{\rm
    e}^{irk}{\tilde \chi}_{k\sigma},
  \label{ft}
\end{equation}
where $k=2\pi n/L$ ($n=-L/2+1, \dots, 0,1,\dots,L/2$) and
$\chi^\dagger_{r\sigma}\equiv (c^{A\dagger}_{r\sigma}\,\,
c^{B\dagger}_{r\sigma})$. Defining the 4-component Nambu spinor
$\Psi^\dagger_k \equiv (c^{\rm A \dagger}_{k\uparrow}, c^{\rm B
  \dagger}_{k\uparrow}, c^{\rm A \phantom\dagger}_{-k\downarrow},
c^{\rm B \phantom\dagger}_{-k\downarrow})$, and with the phase
gradient given by $\phi = \Phi/L$, we obtain
\begin{eqnarray}
  H_{BCS}(\Phi) &=& \sum_k \Psi^\dagger_k {\cal M}_k(\phi) \Psi_k\\
  \nonumber
  &&+L\left ( \frac{1}{U} \sum_{\alpha} (|\Delta^{\alpha}|^2 +
  \theta^{z\,2}_\alpha ) -2\mu -\frac{U}{2}\right ),
  \label{bcsmfthamFTSM}
\end{eqnarray}
which is Eq.(\ref{bcsmfthamFT}). For the general chain depicted in
Fig.\ref{lattice}, the matrix ${\cal M}_k(\phi)$ is
\begin{equation}
  {\cal M}_k(\phi) =
  \begin{pmatrix} {\cal K}(\phi+k) & {\cal D} \\ 
        {\cal D}^* & -{\cal K}^T(\phi-k)
  \end{pmatrix},
  \label{matrixmk}
\end{equation}
where,
\begin{equation}
  {\cal K}(\phi+k) =
  \begin{pmatrix} {\cal K}_{11} -{\bar \mu}^A &&& {\cal K}_{21} \\
        {\cal K}_{12} &&& {\cal K}_{22} -{\bar \mu}^B
  \end{pmatrix},
\end{equation}
with $\bar \mu^\alpha = \mu +U/2 +\theta^z_\alpha$, ${\cal K}_{11}
=t_1{\rm e}^{i(\phi+k)}+t_1^*{\rm e}^{-i(\phi+k)}$, ${\cal K}_{22} =
t_2{\rm e}^{i(\phi+k)}+t_2^*{\rm e}^{-i(\phi+k)}$, ${\cal K}_{12} =
{\cal K}_{21}^*= t_5+t_3{\rm e}^{i(\phi+k)}+t_4{\rm e}^{-i(\phi+k)}$.
The matrix ${\cal D}$ is given by,
\begin{equation}
{\cal D} =
  \begin{pmatrix} \Delta^A & 0 \\ 
        0 & \Delta^B
  \end{pmatrix},
  \label{matrixD}
\end{equation}
The three models we address here are obtained by appropriate choices
of the hopping parameters. The normal ladder is given by
$t_3=t_4=0,\,t_1=t_2=t_5=t$; the flat band Creutz model by
$t_1=t_2^*=it$, $t_3=t_4=t$, $t_5=0$; and the sawtooth model by
$t_1=t_4=0$, $t_2=t$, $t_5=t_3=\sqrt{2}t$.

${\cal M}_k(\phi)$ can be diagonalized giving the ground state energy,
$E_{GS}(\Phi)$,
\begin{eqnarray}
  \nonumber
  E_{GS}(\phi) &=& L\left ( \frac{1}{U} \sum_{\alpha} 
  (|\Delta^{\alpha}|^2 + \theta^{z\,2}_\alpha ) - 2\mu -\frac{U}{2}
  \right ) \\
  && +   \displaystyle\sum_k\left(\lambda_1(k,\phi) +
  \lambda_2(k,\phi)\right), 
    \label{egsgeneral}
\end{eqnarray}
where $\lambda_{1,2}$ are the negative eigenvalues. The mean field
parameters are obtained by minimizing $E_{GS}(\phi)$ with respect to
those parameters or, equivalently, by solving the self consistency
equations,
\begin{equation}
  \Delta^\alpha = U\langle c^\alpha_{i\uparrow} c^\alpha_{i\downarrow}
  \rangle,\hskip 1cm \theta^z_\alpha = U(\langle n^\alpha_i \rangle
  -1)/2,
\end{equation}
with $n^\alpha_i=c^{\alpha\dagger}_{i\uparrow}
c^{\alpha\phantom\dagger}_{i\uparrow} +
c^{\alpha\dagger}_{i\downarrow}c^{\alpha\phantom\dagger}_{i\downarrow}$.

For the normal ladder and Creutz models, $\Delta^A=\Delta^B$ and
$\rho^A=\rho^B$, so that we can use the simple BCS MF with a single
variational parameter, $\Delta$, which is uniform on all sites. The
eigenvalues of ${\cal M}_k(\phi)$ can then be obtained in closed
form. In these cases, the results of the simple BCS MF and the more
extended MF we use here are identical. However, for the sawtooth
lattice with its unequal $A$ and $B$ sites, the two mean field
methods are not equivalent and only the extended method we presented
here gives correct results, especially at strong coupling.

If, instead of using the extended MF method, we use the simple BCS
mean field where only the $\Delta^\alpha$ are MF variational
parameters, $Uc^{\alpha\dagger}_{i\downarrow}
c^{\alpha\dagger}_{i\uparrow} c^{\alpha\phantom\dagger}_{i\uparrow}
c^{\alpha\phantom\dagger}_{i\downarrow} \rightarrow \Delta^{\alpha *}
c^{\alpha\phantom\dagger}_{i\uparrow}
c^{\alpha\phantom\dagger}_{i\downarrow} +
c^{\alpha\dagger}_{i\downarrow}
c^{\alpha\dagger}_{i\uparrow}\Delta^\alpha -|\Delta^\alpha|^2/U$, we
obtain reasonable results only for low $U$ but quantitatively
incorrect results as $U$ increases.
\begin{figure}[ht]
  \includegraphics[width=8.5cm]{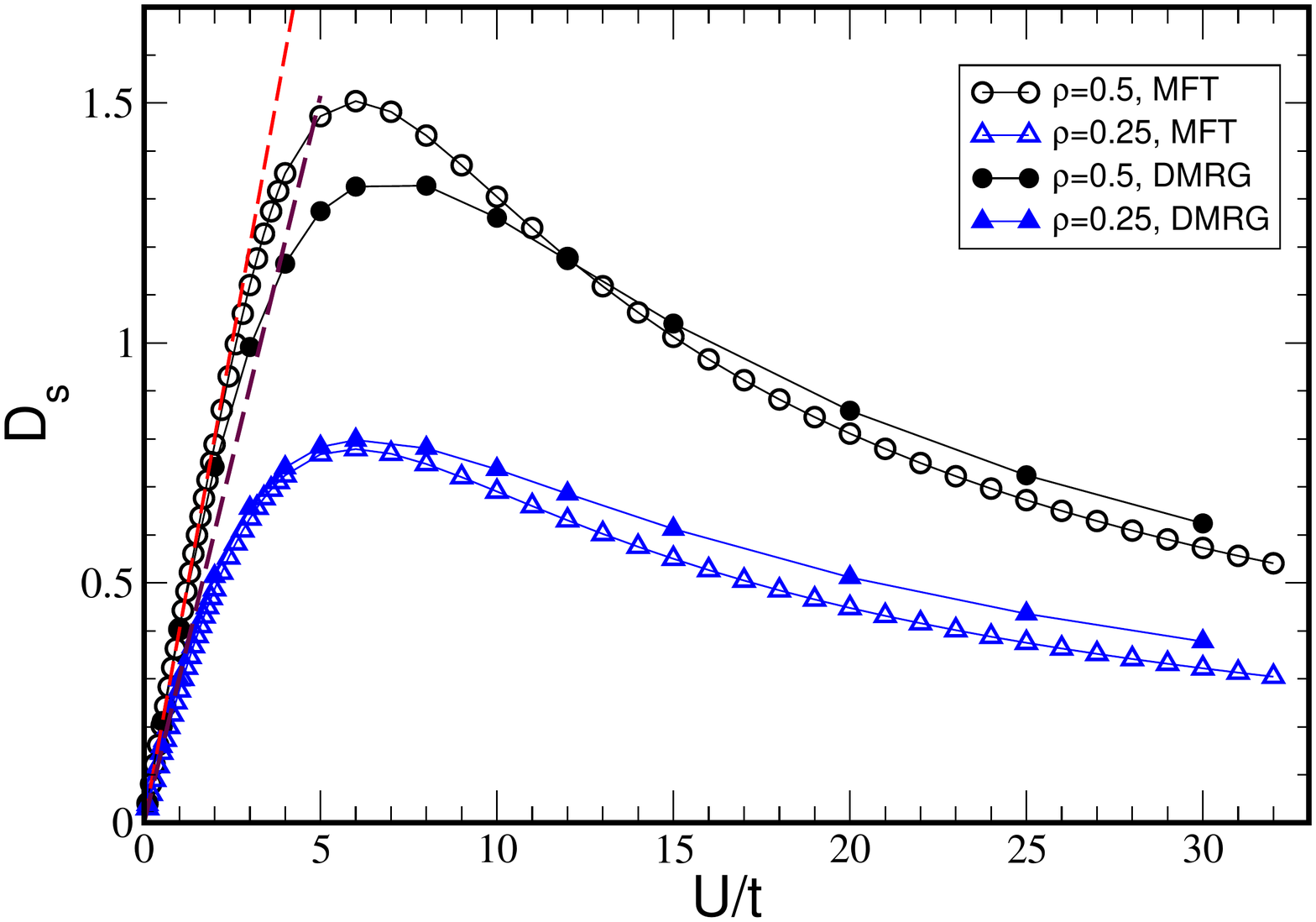}
  \includegraphics[width=8.5cm]{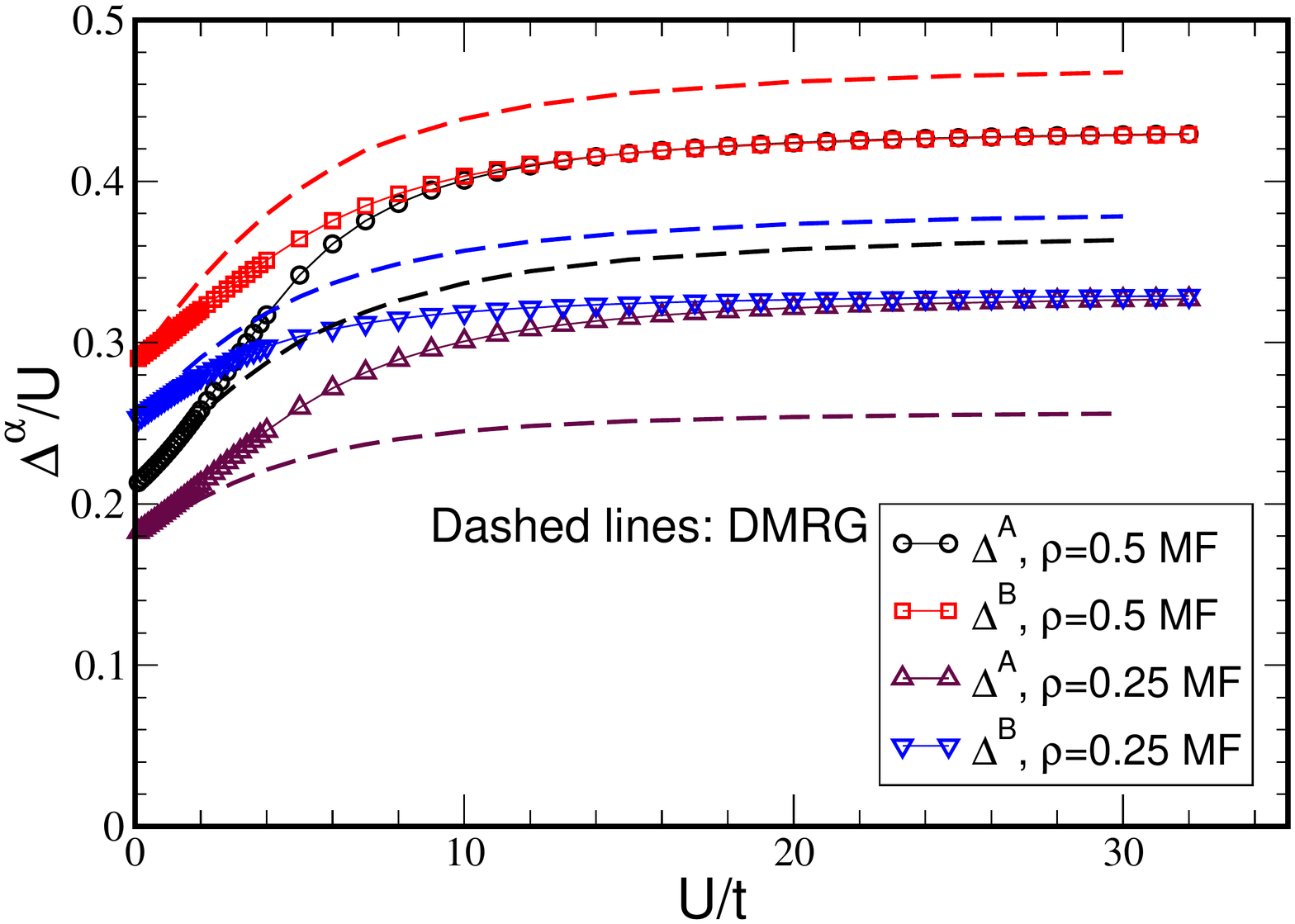}
 \caption{(Color online) Sawtooth lattice. Top panel: $D_s$ versus
   $U/t$ using the simple BCS MF substitution. We find a quite good
   agreement with DMRG for small $U$, but a relatively poor one at
   large U.  Bottom panel: The pairing parameter, $\Delta^\alpha$,
   versus $U/t$. Excellent agreement with DMRG for small $U$. As $U$
   increases, the simple MF results become quantitatively incorrect:
   $\Delta^A \to \Delta^B$ which is not what exact DMRG results
   show. DMRG shows that $\Delta^A$ and $\Delta^B$ remain unequal.}
 \label{sawtoothsimpleBCS}
\end{figure}
The top panel of Fig.\ref{sawtoothsimpleBCS} shows for the sawtooth
lattice $D_s$ versus $U/t$ using the simple BCS MF calculation. We see
that at low $U$, agreement with DMRG is still excellent but for medium
and large values of $U$ the agreement not as good. Compare with
Fig.~\ref{creutzDs} in the main text.  The bottom panel shows, for the
same system, the pairing parameters, $\Delta^A/U$ and $\Delta^B/U$
versus $U/t$ and, again, exhibits good agreement with DMRG for low
values of $U$. However, as $U$ increases, the behavior of
$\Delta^\alpha$ becomes \textit{qualitatively} and
\textit{quantitatively} incorrect: The simple MF shows that, at large
$U$, $\Delta^A=\Delta^B$, which the exact DMRG results show never
happens (compare with Fig.~\ref{deltasvsUsaw}(a)).  The same behavior
is observed for $\rho^A$ and $\rho^B$ (not shown). It is therefore
crucial to use the correct mean field decoupling in order to obtain
qualitatively correct (and quantitatively accurate) results.

\section{Superfluid weight for a flat band system with inequivalent sites}
\label{appendix:ds}

\subsection{General situation}

In a mean-field approach, the superfluid weight reads:
\begin{equation}
 D_s=\pi \left.\frac{d^2\epsilon_{GS}\left(\Delta^{\alpha}(\phi),
   \Delta^{\alpha*}(\phi), \phi\right)}{d^2\phi}\right|_{\phi=0}
\end{equation}
where, $\phi$ is the phase gradient and, $\epsilon_{GS}=E_{GS}/L$ is
the ground state energy per unit cell, see Eq.~\eqref{egsgeneral},
\begin{equation}
\epsilon_{GS}(\Delta^{\alpha},\Delta^{\alpha*},\phi)=
\frac{1}{U}\sum_{\alpha}|\Delta^{\alpha}|^2+\frac{1}{L}\sum_{k,n}
\lambda_n(k,\Delta^{\alpha}, \Delta^{\alpha*},\phi),
\end{equation}
where we have made explicit the dependence on $\Delta^{\alpha}$. For
simplification, we have omitted (i) the $\theta^{\alpha}$ terms which
do not play an important role at low $U$, and (ii) the explicit
dependence on the chemical potential $\mu$ since, even for
inequivalent sites, it does not give any additional contribution to
$D_s$.

Differentiating $\epsilon_{GS}$ twice with respect to $\phi$, and
using the fact that we have $\frac{\partial \epsilon_{GS}}{\partial
  \Delta^{\alpha}}=\frac{\partial
  \epsilon_{GS}}{\partial\Delta^{\alpha*}}=0$ along the mean-field
solution, $\Delta^{\alpha}=\Delta_{mf}^{\alpha}(\phi)$, one obtains
\begin{align}
 \frac{D_s}{\pi}=\frac{1}{L}\sum_{k,n}&\left(\frac{\partial^2
   \lambda_n}{\partial \phi^2}+ \sum_{\alpha}\frac{\partial^2
   \lambda_n}{\partial \Delta^{\alpha}\partial\phi}
 \frac{d\Delta_{mf}^{\alpha}}{d\phi}\right.\\ &\left.+
 \sum_{\alpha}\frac{\partial^2 \lambda_n}{\partial
   \Delta^{\alpha*}\partial\phi}
 \frac{d\Delta_{mf}^{\alpha*}}{d\phi}\right),
\end{align} 
where all derivatives are computed at $\phi=0$ and
$\Delta^{\alpha}=\Delta_{mf}^{\alpha}(\phi=0)$.  For systems which are
invariant under the time-reversal symmetry, and with equivalent sites,
the $U(1)$ symmetry allows us to have all $\Delta^{\alpha*}$ to be
real, so that $\frac{d\Delta_{mf}^{\alpha*}}{d\phi}=0$ at $\phi=0$. In
this situation, the only contribution to $D_s$ is the first term
which, in the $U\rightarrow 0$ limit, can be shown to be proportional
to the quantum metric~\cite{peotta15}.

Generally, in the $U\rightarrow 0$ limit and projecting on the flat
band, one can approximate the chemical potential by
$\mu\approx\epsilon_{FB}+a U$, and the mean field parameters by
$\Delta^{\alpha}\approx U \tilde{\Delta}^{\alpha}$, thereby allowing
us to factor out the $U$ dependence of the ground state energy:
 \begin{align}
 \epsilon_{GS}(\tilde{\Delta}^{\alpha},\tilde{\Delta}^{\alpha*},\phi)&
 =-\frac{U}{L}\sum_k d_k(\tilde{\Delta}^{\alpha},\tilde{\Delta}^{\alpha*},\phi)
 \end{align}
with
\begin{align}
  \label{appBdk}
  d_k(\tilde{\Delta}^{\alpha},\tilde{\Delta}^{\alpha*},\phi)&=
  \sqrt{a^2+|\tilde{b}_{k}(\tilde{\Delta}^{\alpha},\phi)|^2}\\
  \tilde{b}_{k}(\tilde{\Delta}^{\alpha},\phi)&=
  \sum_{\alpha}P^*_{\alpha}(-k+\phi)\tilde{\Delta}^{\alpha}
  P^*_{\alpha}(k+\phi),
  \label{appbtildebk}
 \end{align}
where $P_{\alpha}(k)$ are the site components of the Bloch vector of
the flat band, i.e. the eigenvector of the matrix $\mathcal{K}(k)$.
The mean-field parameters and $a$ fulfill the following
self-consistent equations:
\begin{align}
 \rho&=\frac{1}{2}+\frac{a}{2L} \sum_k
 \frac{1}{d_k(\tilde{\Delta}^{\alpha}, \tilde{\Delta}^{\alpha*}, \phi)}\\
 \tilde{\Delta}^{\alpha}&=\frac{1}{2L}\sum_k
 \frac{P_{\alpha}(-k+\phi) P_{\alpha}(k+\phi)
   \tilde{b}_{k}(\tilde{\Delta}^{\alpha},\phi)}{d_k(\tilde{\Delta}^{\alpha},
   \tilde{\Delta}^{\alpha*},\phi)}\label{eq:mf_self_consistent}, 
 \end{align} 
 and the superfluid weight reads:
 \begin{align}  
  D_s=-U\pi&\left(\frac{1}{L}\sum_k \left.\frac{\partial^2
    d_k}{\partial
    \phi^2}\right|_{\tilde{\Delta}_{mf}^{\alpha}(0),\phi=0}\right.\\
  \nonumber
  &+\frac{1}{L}\sum_k \sum_{\alpha} 
  \left.\frac{\partial^2 d_k}{\partial
    \phi\partial\tilde{\Delta}^{\alpha}}
  \right|_{\tilde{\Delta}_{mf}^{\alpha}(0),\phi=0}
  \left.\frac{d\tilde{\Delta}_{mf}^{\alpha}}{d\phi}\right|_{\phi=0}\\
   &+\left.\frac{1}{L}\sum_k \sum_{\alpha} 
   \left.\frac{\partial^2 d_k}{\partial
     \phi\partial\tilde{\Delta}^{\alpha*}}\right|_{\tilde{\Delta}_{mf}^{\alpha}(0),\phi=0} 
   \left.\frac{d\tilde{\Delta}_{mf}^{\alpha*}}{d\phi}\right|_{\phi=0}\right).
   \nonumber
\end{align}
Differentiating the self-consistent
equation~\eqref{eq:mf_self_consistent} with respect to $\phi$ allows
us to find a set of coupled linear equations fulfilled by all
$\left.\frac{d\tilde{\Delta}^{\alpha}_{mf}}{d\phi}\right|_{\phi=0}$. For
time-reversal invariant system, we can show that all these quantities
are purely imaginary, and the set of linear equations reads formally
$O_{\alpha\alpha'}X_{\alpha'}=Y_{\alpha}$, with
$X_{\alpha}=\left.\frac{d\tilde{\Delta}^{\alpha}_{mf}}{d\phi}\right|_{\phi=0}$
and
\begin{align}
 O_{\alpha\alpha'}&=\delta_{\alpha\alpha'}-\frac{1}{2L}\sum_k
 \frac{P_{\alpha}(k) P^*_{\alpha}(k)P_{\alpha'}(k) P^*_{\alpha'}(k)
 }{d_k}\\ 
 Y_{\alpha}&=\frac{1}{2L}\sum_k\frac{b_k}{d_k}\nonumber
 \left(P^*_{\alpha}(k) \partial_k P_{\alpha}(k)- \partial_k
 P^*_{\alpha}(k) P_{\alpha}(k)\right)\\ 
 &+\frac{1}{2L}\sum_k \frac{P_{\alpha}(k)
   P^*_{\alpha}(k)}{d_k}\partial_{\phi} b_k.
\end{align}
Note that the matrix $O$ is singular. Indeed, one can see that for
$X_{\alpha}=\tilde{\Delta}_{mf}^{\alpha}$ at $\phi=0$, the self-consistent
equations lead to
$\sum_{\alpha'}O_{\alpha\alpha'}\tilde{\Delta}_{mf}^{\alpha}=0$, which
is just the $U(1)$ symmetry: if one adds a global phase to all MF
parameters, it does not change the GS, i.e. the self-consistent
equations are still fulfilled.

\subsection{Two-band lattices}

In the specific case of a two-band system, like the sawtooth lattice,
the $2\times2$ matrix $\mathcal{K}(k)$ can be written formally as
follows,
\begin{align}
  \mathcal{K}(k)&=e_0(k)\openone + \vec{e}(k)\cdot\vec{\sigma}
  \nonumber \\ 
 \vec{e}(k)&=|e(k)| \left(\sin{\vartheta_k} \cos{\varphi_k},
 \sin{\vartheta_k}  \sin{\varphi_k}, \cos{\vartheta_k}\right),
\end{align}
where ${\vec \sigma}$ are the Pauli matrices. Assuming that the flat
band corresponds to the lower band, the Bloch eigenvector reads
\begin{equation}
P_{A}(k)=-\sin{\frac{\vartheta_k}{2}}e^{-i\varphi_k/2}\qquad
P_{B}(k)=\cos{\frac{\vartheta_k}{2}}e^{i\varphi_k/2}.
\end{equation}
Let us introduce the notations:
\begin{equation}
 \tilde{\Delta}^A=\tilde{\Delta}(1-\delta)\text{ and }
 \tilde{\Delta}^B=\tilde{\Delta}(1+\delta), 
\end{equation}
 and, using Eqs.(\ref{appBdk},\ref{appbtildebk}), we find at $\phi=0$
\begin{align}
 \tilde{b}_k&= \tilde{\Delta}
 \biggl(1+\delta\cos{\vartheta_k}\biggr)\\
 d_k&=\sqrt{a^2+\tilde{\Delta}^2\left(1+\delta\cos{\vartheta_k}\right)^2}
\end{align}
fulfilling the following equations:
\begin{align}
 \rho-\frac{1}{2}&=\frac{a}{2L}\sum_k\frac{1}{d_k}\\ 1&=\frac{1}{4L}\sum_k
 \frac{1+\delta\cos{\vartheta_k}}{d_k}\\ \delta&=\frac{1}{4L}\sum_k
 \frac{\cos{\vartheta_k}\left(1 + \delta\cos{\vartheta_k}\right)}{d_k}
 \end{align}
 
 After some straighforward computations, one can show that the
 superfluid weight reads:
\begin{align}
\label{eq:sf_full}
 D_s=U\tilde{\Delta}^2\Biggl(&\frac{2\pi}{L}\sum_k
 \frac{\mathcal{B}_k}{d_k}\nonumber\\ +&\delta\frac{2\pi}{L}
 \sum_k\frac{\left(\sin{\vartheta_k}\partial^2_k\vartheta_k +
   \cos{\vartheta_k}(\partial_k\vartheta_k)^2
   \right)}{2d_k}\nonumber\\
 +&\delta^2\frac{2\pi}{L}\sum_k \frac{\left(\sin{\vartheta_k}
   \cos{\vartheta_k} \partial^2_k\vartheta_k-
   \sin^2{\vartheta_k}(\partial_k\varphi_k)^2 
   \right)}{2d_k}\nonumber\\ &+\gamma(1-\delta^2)
 \frac{2\pi}{L}\sum_k\frac{(1-\cos^2{\vartheta_k})
   \partial_k\varphi_k}{2d_k}\Biggr), 
\end{align}
where 
\begin{align}
 \mathcal{B}_k&=2\sum_{\alpha}\partial_k P^*_{\alpha}\partial_k
 P_{\alpha}-2|\sum_{\alpha} \partial_k
 P^*_{\alpha}P_{\alpha}|^2\nonumber\\ 
 &=\frac{1}{2}\left( (\partial_k\vartheta_k)^2+\sin^2{\vartheta_k}
 (\partial_k\varphi_k)^2\right),
\end{align}
is the quantum geometric tensor~\cite{tovmasyan16} in 1D. For systems
where the sublattices are equivalent, i.e. $\delta=0$, one can show
that $d_k=1/4$, such that
\begin{align}
 \frac{E_{GS}}{L}&=-\frac{U}{4}\\
 \mu&=\epsilon_{FB}+\frac{U}{2}(\rho-\frac{1}{2})\\
 \tilde{\Delta}&=\frac{1}{2}\sqrt{\rho(1-\rho)}\\
 D_s&=2\pi U\frac{\rho(1-\rho)}{4}4\frac{1}{L} \sum_k
 \mathcal{B}_k=2\pi U\rho(1-\rho)\mathcal{Q},
\end{align}
where
\begin{equation}
  \mathcal{Q}=\frac{1}{2\pi}\int_{BZ}\mathcal{B}_k
  \label{appBquantmetric}
\end{equation}
is the quantum metric. This recovers the results of
ref~\cite{tovmasyan16}.

The last term in Eq.~\eqref{eq:sf_full} is the contribution from the
first derivative of the meanfield parameters, with
\begin{align}
 \gamma&=-i\frac{1}{\tilde{\Delta}^{A}_{mf}(\phi=0)}
 \left. \frac{d\tilde{\Delta}^{A}_{mf}}{d \phi}\right|_{\phi=0} 
 =i\frac{1}{\tilde{\Delta}^{B}_{mf}(\phi=0)}
 \left. \frac{d\tilde{\Delta}^{B}_{mf}}{d\phi}\right|_{\phi=0}\nonumber\\
 &=-\frac{\frac{1}{L}\sum_k\frac{\partial_k\varphi_k(1 -
     \cos^2{\vartheta_k})}{d_k}} {\frac{1}{L}\sum_k\frac{1 -
     \cos^2{\vartheta_k}}{d_k}} 
\end{align}

\begin{figure}[h]
 \includegraphics[width=0.45\textwidth]{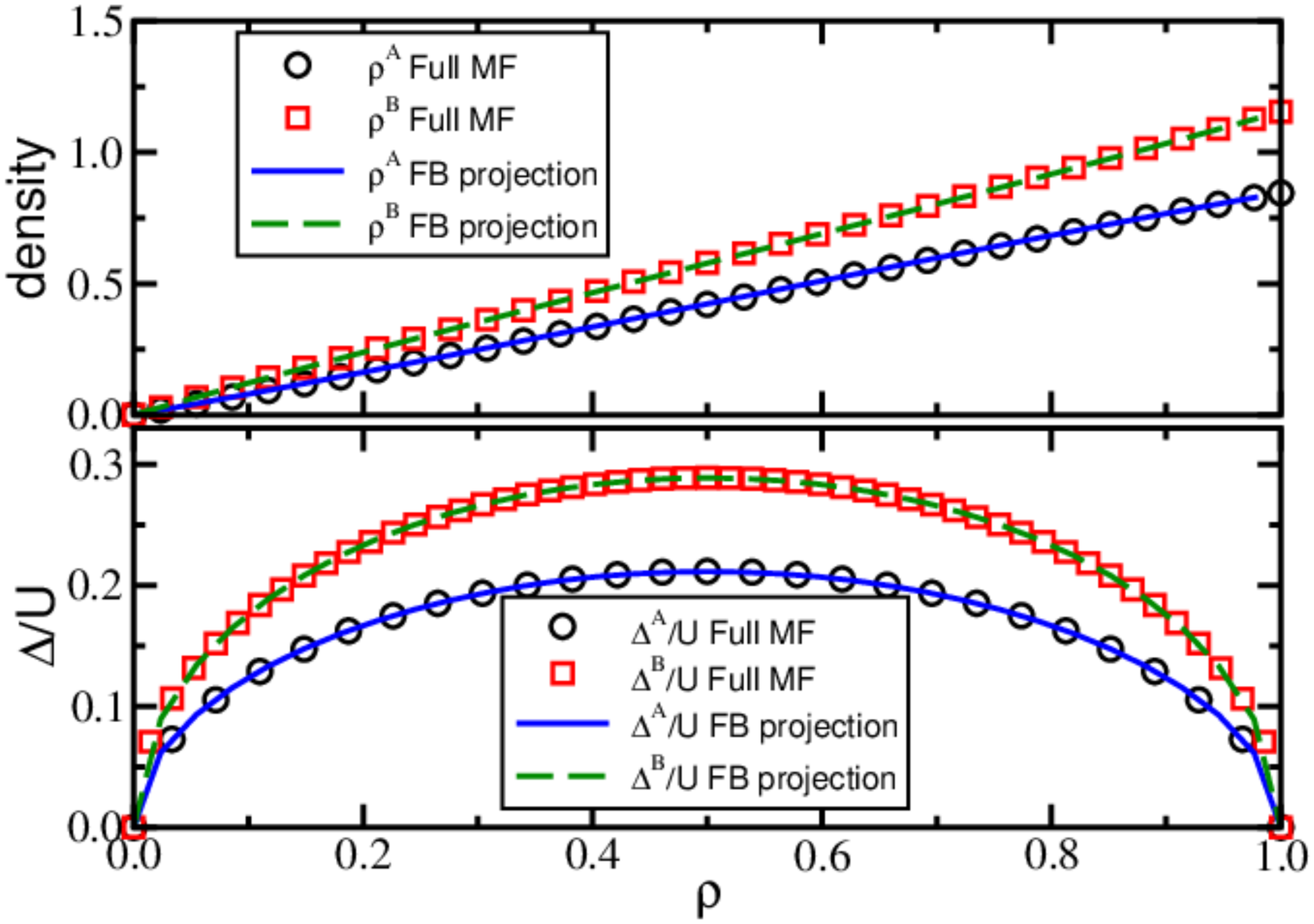}
 \caption{\label{fig:mfres} (Color online) Mean field parameters
   $\Delta^{\alpha}/U$ and densities $\rho^{\alpha}$ vs the total
   density $\rho$, extrapolated to $U=0$.  Lines: our projection on
   the flat band; symbols: Full multi-band mean field.}
\end{figure}

For the specific case of the sawtooth lattice, one has
\begin{equation}
 \vec{e}(k)=\left(\sqrt{2}t(1 + \cos{k}), -\sqrt{2}t\sin{k},
 -t\cos{k}\right),
\end{equation}
such that
\begin{equation}
  \cos{\vartheta_k}=-\frac{\cos{k}}{2+\cos{k}}\text{ and }
  \varphi_k=-\frac{k}{2}.
\end{equation}
This allows us to compute all quantities $a$, $\delta$ and
$\tilde{\Delta}$ as functions of the total density $\rho$. It turns
out that
\begin{align}
 a&\approx\frac{1}{2}\left(\rho-\frac{1}{2}\right),\\
 \tilde{\Delta}&\approx\frac{1}{2}\sqrt{\rho(1-\rho)},\\
 \delta&\approx \delta_0+\frac{1}{8}\left(\rho-\frac{1}{2}\right)^2,
\end{align}
where $\delta_0=0.154701$ is the exact value of $\delta$ at
$\rho=\frac{1}{2}$.  The approximate solution gives
\begin{align}
 \tilde{\Delta}^A&=\frac{1}{2}\sqrt{\rho(1-\rho)}\left( 1-
 (\delta_0+\frac{1}{8}\left(\rho - \frac{1}{2}\right)^2\right),\\
 \tilde{\Delta}^B&=\frac{1}{2}\sqrt{\rho( 1- \rho)}\left(1 +
 (\delta_0+\frac{1}{8}\left(\rho - \frac{1}{2}\right)^2\right),
\end{align}
and the sub-lattice densities, 
\begin{equation}
 \rho^{\alpha}=\rho\mp\frac{1}{4\pi} \int_{BZ}\, dk
 \cos{\vartheta_k}\left(\frac{a}{d_k} + 1\right),
\end{equation}
where $\mp$ corresponds to $A$, resp. $B$. All results are displayed
in Fig.~\ref{fig:mfres}. As one can, see the agreement between the
full multi-band MF and the flat band projected mean field (in the $U\to
0$ limit) is excellent for all values of the total density $\rho$.

Along the same lines, one can compute the single particle Green
functions:
\begin{equation}
 G^{\alpha\alpha}_{\sigma}(r)=-\frac{1}{8\pi}\int_{BZ}\, dk
 \left(1\mp\cos\vartheta_k\right)\left(\frac{a}{d_k} + 1\right)e^{ikr},
\end{equation}
and the two-point correlation function 
\begin{equation}
 \langle c^{\alpha\dagger}_{j+r\uparrow}c^{\alpha\dagger}_{j\downarrow}\rangle
 =\frac{\tilde{\Delta}}{8\pi}\int_{BZ}\, dk
 \left(1\mp\cos\vartheta_k\right)\frac{1+\delta\cos\vartheta_k}{d_k}e^{ikr}.
\end{equation}
In the large distance limit, one can show that $\langle
c^{\alpha\dagger}_{j+r\uparrow}c^{\alpha\dagger}_{j\downarrow}\rangle$
decays exponentially over a length scale that is associated to the
size of the BCS pairs. It turns out that for $\rho=0.5$, one has $a=0$
and $d_k=|1+\delta\cos\vartheta_k|$, such that the single particle
Green function and the two-point correlation function are given by the
same expression (up to a sign):
\begin{equation}
\frac{1}{8\pi}\int_{BZ}\, dk \left(1\mp\cos\vartheta_k\right)e^{ikr},
\end{equation}
leading an exponential decay at large distance $G^{\alpha\alpha}_{\sigma}(r)
\propto \exp{-r/\xi}$, with $\xi\approx0.759$.

\bibliography{flatbandrefs}

\end{document}